\documentclass[aps,prb,preprint,superscriptaddress,showpacs]{revtex4-1}

\usepackage{graphicx} 
\usepackage{textcomp} 
\usepackage{hyperref} 
\usepackage{ifthen} 
\usepackage{xifthen} 
\usepackage{color} 
\usepackage{soul} 
\usepackage{amsmath} 
\usepackage{lineno}
\usepackage[english]{babel}
\usepackage{ucs}
\usepackage[utf8x]{inputenc}
\usepackage{amssymb}
\usepackage{amsfonts}
\usepackage{bm}
\usepackage{bbm}

\newcommand{\abs}[1]{\left|#1\right|}

\newcommand{\fig}[2][]{%
\ifthenelse{\isempty{#1}}
{Fig.~\ref{#2}}
{Fig.~\ref{#2}(#1)}
}
\newcommand{\Fig}[2][]{%
\ifthenelse{\isempty{#1}}
{Figure~\ref{#2}}
{Figure~\ref{#2}(#1)}
}

\begin{document}


\title{Parametric oscillation, frequency mixing and injection locking of strongly coupled nanomechanical resonator modes}



\author{Maximilian\,J. Seitner}
\email{maximilian.seitner@uni-konstanz.de}
\affiliation{Department of Physics, University of Konstanz, 78457 Konstanz, Germany}
\author{Mehdi Abdi}
\affiliation{Department of Physics, Technische Universit\"at M\"unchen, 85748 Garching, Germany}
\affiliation{Institute for Theoretical Physics, Ulm University, 89081 Ulm, Germany}
\author{Alessandro Ridolfo}
\affiliation{Dipartimento di Scienze Matematiche e Informatiche, Scienze Fisiche e Scienze della Terra (MIFT), Universita di Messina, 98166 Messina, Italy}
\author{Michael\,J. Hartmann}
\affiliation{Institute of Photonics and Quantum Sciences, Heriot-Watt University, EH14 4AS Edinburgh, United Kingdom}
\author{Eva\,M. Weig}
\affiliation{Department of Physics, University of Konstanz, 78457 Konstanz, Germany}


\begin{abstract}
We study locking phenomena of two strongly coupled, high-quality factor nanomechanical resonator modes subject to a common parametric drive at a single drive frequency. By controlled dielectric gradient forces we tune the resonance frequencies of the flexural in-plane and out-of-plane oscillation of the high stress silicon nitride string through their mutual avoided crossing. For the case of the strong common parametric drive signal-idler generation via parametric oscillation is observed, analogously to the framework of nonlinear optical effects in an optical parametric oscillator. Frequency tuning of the signal and idler resonances is demonstrated. When the resonance frequencies of signal and idler get closer to each other, partial injection locking, injection pulling and complete injection locking to half of the drive frequency occurs depending on the pump strength. Furthermore, satellite resonances, symmetrically off-set from signal and idler by their beat-note, are observed which can be attributed to degenerate four-wave-mixing in the highly nonlinear mechanical oscillations.
\end{abstract}

\maketitle

\begin{figure}[!htb]
\includegraphics{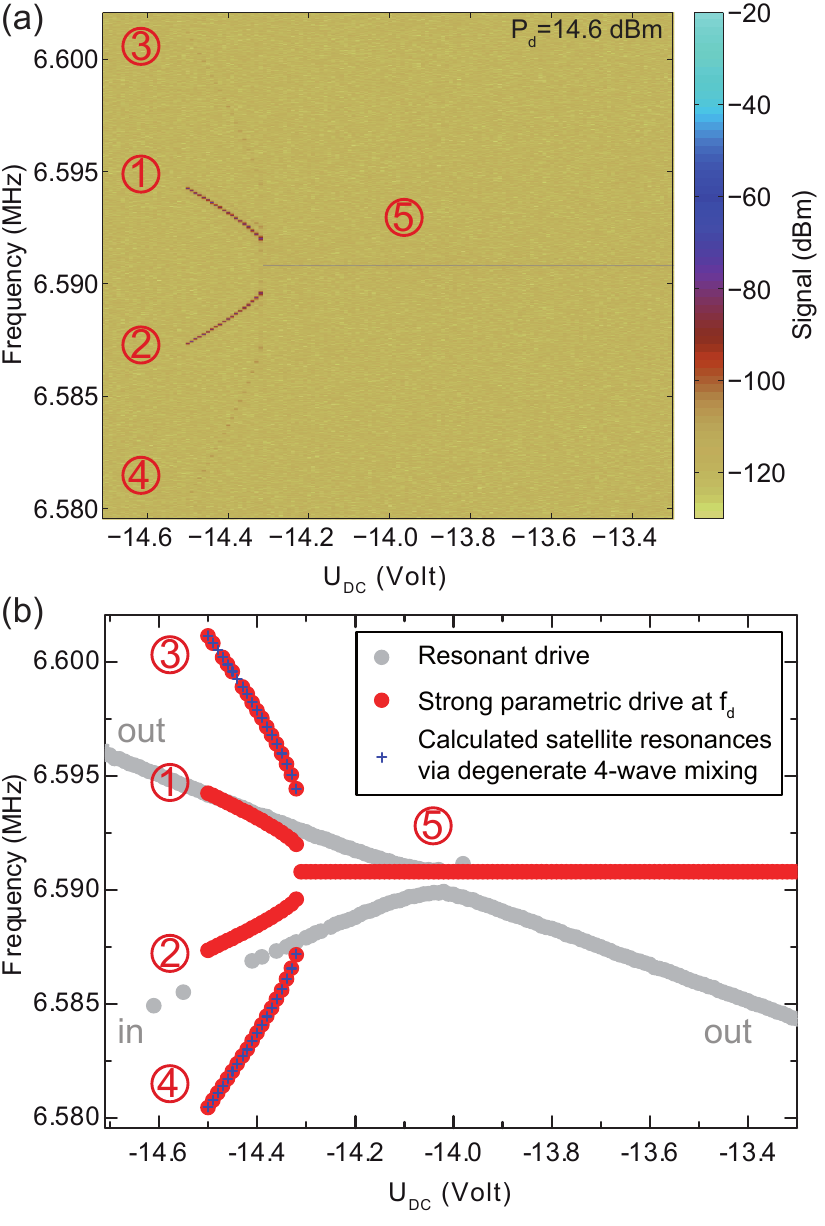}
\caption{\label{fig1} (a) Color-coded frequency spectra versus DC voltage for a strong parametric drive power $P_\mathrm{d}=14.6$\,dBm at $f_\mathrm{d}=13.18163$\,MHz. Parametric oscillation resonances ($f_1$, $f_2$), satellites from degenerate four-wave-mixing ($f_3$, $f_4$) and complete injection locking ($f_5$) to half of the drive frequency are observed. (b) Extracted peak frequencies of the above spectra (red dots) and calculated satellite resonances from Eq.~\eqref{eq:4wm}(blue daggers) exhibit excellent agreement. Gray dots display the avoided crossing of the two modes under weak resonant drive with mode splitting of $0.97$\,kHz. Some gray dots are missing for the in-plane mode due to bad detection efficiency of this mode.}
\end{figure}

The development of lasers and masers opened up the path to the realm of controlled nonlinear physics. If a material with a nonlinear susceptibility is strongly pumped above a certain threshold by a coherent drive source, its response exhibits characteristic nonlinear effects. The Kerr effect, second-harmonic generation, spontaneous parametric down-conversion or four-wave-mixing are just some prominent examples for the great variety of nonlinear physical effects\,\cite{Saleh_Teich}. Installing the nonlinear medium inside a resonant cavity leads to the concept of parametric oscillators, in particular the optical parametric oscillator\,\cite{Yariv} (OPO). During the past decades, this model system has been exploited to study a variety of parametric effects including injection locking\,\cite{1973_Buczek_IEEE}, where an oscillator is strongly pumped at a frequency close to its natural resonance frequency. If the amplitude of the pump exceeds a threshold value, the oscillator will lock to the pump in frequency and/or phase. This phenomenon was originally proposed by Adler\,\cite{1973_Adler_IEEE} and has been studied in a variety of systems, for example, semiconductor lasers\,\cite{1994_Annovazzi_IEEE,1998_Annovazzi_IEEE,2001_Liu_IEEE,2003_Liu_IEEE}, electrical oscillators\,\cite{1966_Stover_IEEE,1989_Chang_IEEE,2004_Razavi_IEEE,2007_Mirzaei_IEEE}, organ pipes\,\cite{2006_Abel_JAS,2009_Abel_PRL} and micromasers\,\cite{2015_Liu_PRA} amongst others.\\
For the case of simultaneous frequency and phase locking in combination with the reduction of phase noise, synchronization of two or more oscillators to a reference oscillator or to each other\,\cite{1973_Adler_IEEE} is observed.\\
Recently, the study of parametric oscillation\,\cite{Rugar1991,1998_Turner_Nature,2016_Mathew_NatNano,2016_Leuch_PRL}, thermal-noise two-mode squeezing\,\cite{2014_Mahboob_PRL,2015_Patil_PRL,2016_Pontin_PRL,2016_Mahboob_NJP} and synchronization phenomena has been transferred to mechanical microresonators\,\cite{2004_Cross_PRL,2011_Heinrich_PRL,2011_Thevenin_PRL,2012_Holmes_PRE,2014_DelHaye_PRL,2014_Barois_NJP,2016_Li_OptExp} and the field of cavity opto-/electro-mechanics. In the latter case, special attention needs to be paid to distinguish between demonstration of real synchronization as originally defined above\,\cite{2013_Bagheri_PRL,2013_Deepak_PRL,2014_Matheny_PRL,2014_Gieseler_PRL,2015_Shah_PRL,2015_Zhang_PRL} and mixing and locking phenomena of oscillators\,\cite{Mohanty,2008_Zadeh_APL,2012_Huang_APL,2012_Zhang_PRL,2015_Antonio_PRL,2016_Gil-Santos_arxiv}.\\
In this work, we demonstrate a set of characteristic nonlinear physical effects in a nanoelectromechanical system. In particular, we observe frequency mixing products which are analogous to second-order and third-order nonlinear effects in nonlinear optical systems. The nanoelectromechanical system consists of two strongly coupled high quality factor nanomechanical modes subject to a joint strong parametric pump. Depending on the frequency difference of the two modes and the strength of the parametric drive, we observe signal-idler generation via parametric oscillation, satellite resonances induced by degenerate four-wave-mixing (D4WM), partial injection locking including injection pulling effects, and injection locking over a large frequency range. The effects are investigated by measuring the frequency spectra of the two coupled nanomechanical resonator modes near the avoided energy level crossing. Since we are not able to measure the phase relation between the two oscillations, we deliberately avoid the term \textit{synchronization} throughout the remainder of the manuscript.\\
The employed nanoelectromechanical system (cf. Supplemental Material\,\cite{SI_lock}) consists of the flexural in-plane (in) and out-of-plane (out) mode of a $55$\,\textmu m long, $270$\,nm wide and $100$\,nm thick, doubly clamped silicon nitride string resonator operated in vacuum at room temperature, which is flanked by two adjacent gold electrodes. A DC voltage applied to the electrodes enables dielectric tuning ($\propto U_\mathrm{DC}^2$) of the resonance frequencies ($f_\mathrm{0}\approx 6.5$\,MHz) via electric gradient fields\,\cite{2012_Rieger_APL}. When the two modes are tuned into resonance they hybridize into normal modes, i.e., linear combinations of in-plane and out-of-plane, and exhibit a pronounced avoided crossing, reflecting the strong coupling of the nanomechanical two-mode system  provided by the inhomogeneous electric field\,\cite{2012_Faust_PRL,2013_Faust_NatPhys,2016_Seitner_PRB}. The mechanical modes exhibit a high mechanical quality factor ($Q_\mathrm{0}\approx 500,000$) which reduces quadratically with the applied DC voltage\,\cite{2012_Rieger_APL}. A microwave cavity connected via the same two electrodes is employed for signal transduction and heterodyne read-out\,\cite{2012_Faust_NatComm}.\\
We investigate the dynamics of the two coupled modes in the vicinity of the avoided crossing under a strong common parametric drive tone for drive powers between $P_\mathrm{d}=-1.5$\,dBm and $P_\mathrm{d}=15$\,dBm by applying the following measurement scheme:\newline
We initialize the system to the left-hand side of the avoided crossing by choosing an appropriate DC voltage (cf. Fig.\,\ref{fig1}). At this point, the two modes are thermally excited at their resonance frequencies by the thermal noise of the room temperature environment. Note that the thermal amplitudes of the two modes are too small to be resolved in the frequency spectra discussed in the following. A non-resonant strong sinusoidal drive tone at $f_\mathrm{d}$ is continuously applied to the system. This parametric drive tone modulating the resonant frequencies at approximately twice the eigenfrequencies of the out-of-plane and in-plane mode, $f_\mathrm{d}\approx 2f_\mathrm{out}\approx 2f_\mathrm{in}$, is also applied to the electrodes as a corresponding RF voltage $U_\mathrm{d}\propto \sqrt{P_\mathrm{d}}\sin(2\pi f_\mathrm{d}t)$. Note that this RF voltage does not effectuate a resonant driving for the lack of a mechanical mode at this frequency and hence the system can be viewed as a pair of non-degenerate parametric oscillators\,\cite{2001_Olkhovets_IEEE,2014_Mahboob_PRL,2016_Mahboob_NJP}. By changing the DC voltage, the two thermally excited resonances are tuned through the avoided crossing where we record the mechanical frequency spectra for each particular $U_\mathrm{DC}$.\\
Prior to each measurement, the avoided crossing of the two modes is recorded in the weakly driven regime by tuning the DC voltage, where the mechanical modes are resonantly excited by a weak 10\,MHz bandwidth noise drive. The coupling strength is determined from the avoided crossing data and the center frequency $f_\mathrm{c}$ is chosen as a frequency in between the two modes at the avoided crossing frequency splitting. Note that the center frequency does not need to be chosen as the mean of the two respective mode frequencies. The frequency of the strong parametric sinusoidal drive tone $f_\mathrm{d}$ corresponds to twice the center frequency $f_\mathrm{d}=2 f_\mathrm{c}$.\\
Figure\,\ref{fig1}\,(a) depicts the color-coded frequency spectra for a strong parametric drive of $P_\mathrm{d}=14.6$\,dBm. In Fig.\,\ref{fig1}\,(b) the extracted resonance peaks of the frequency spectra are displayed as red dots. The gray dots visualize the resonantly driven response of the system, i.e., the avoided crossing of in-plane ($f_\mathrm{in}$) and out-of-plane ($f_\mathrm{out}$) resonances with a level splitting of $0.97$\,kHz. At a voltage of $U_\mathrm{DC}=-14.5$\,V two sharp frequency peaks at $f_1\approx f_\mathrm{out}(-14.5$\,V) and $f_2\neq f_\mathrm{in}(-14.5$\,V) suddenly appear. From the extracted frequencies we find
\begin{equation}
\label{eq:opa}
f_1 + f_2 = f_\mathrm{d} = 2 f_\mathrm{c},
\end{equation}
which mimics the signal-idler generation process in optical parametric oscillation\,\cite{Yariv}. The strong pump at $f_\mathrm{d}$ parametrically amplifies (non-degenerate amplification above the parametric instability threshold) the thermally excited resonance of the out-of-plane mode (signal) if its resonance frequency gets close enough to half of the pump frequency. Simultaneously, an idler resonance at frequency $f_2 = f_\mathrm{d} - f_1$ is created due to the coupling of the two modes. The fact that the idler frequency $f_2$ differs from the resonance frequency of the natural in-plane mode clearly establishes the analogy to the process of signal-idler generation via optical parametric oscillation in a doubly resonant cavity, which is a second-order nonlinear optical effect. The in-plane mode is forced to oscillate at a frequency determined from the frequency-matching condition (Eq.~\eqref{eq:opa}) which is analogous to the phase-matching condition in optical parametric oscillation\,\cite{Yariv}. In the present experiment, the optical cavity is replaced by the two flexural modes of the nanomechanical string resonator and the nonlinear optical medium corresponds to the highly nonlinear response of the resonant modes under parametric oscillation. The mechanical nonlinearity is intrinsic to the material and becomes apparent because of the high mechanical quality factors and the corresponding large vibration amplitudes of the in-plane and out-of-plane mode. After the parametric instability threshold, we are able to tune the two sharp parametric resonances towards each other in frequency as a function of the DC voltage. Note that the tuning behaviour slightly deviates from the frequency tuning curves of the natural resonator modes subject to a weak resonant drive, even for the signal parametric resonance ($f_1$) and the natural out-of-plane mode ($f_\mathrm{out}$). We attribute this to the high amplitudes of the parametric resonances compared to the amplitudes of the two modes when linearly driven on resonance. The electric field gradient of the inhomogeneous electric field can only be linearized as a function of the DC voltage in the limit of small deflections of the string resonator\,\cite{2012_Rieger_APL} which gives rise to the quadratic tuning of $f_\mathrm{out}$, whereas a modified tuning curve is expected for $f_1$.\\
Additionally, we observe two satellite resonances which are off-set from the parametric resonances by their frequency difference and hence correspond to their beat-note. Hence, we interpret the appearance of the two satellites to the third-order nonlinear process of degenerate four-wave-mixing\,\cite{Saleh_Teich} (D4WM) since the frequencies of the satellites can be expressed as mixing products of the present oscillation frequencies:
\begin{eqnarray}
\label{eq:4wm}
f_3 &= 2 f_1 - f_2
\qquad
f_4 &= 2 f_2 - f_1
\end{eqnarray}
In Figure\,\ref{fig1}\,(b) the calculated frequencies following Eq.~\eqref{eq:4wm} are displayed as blue daggers and perfectly match the satellite frequencies extracted from the experimental data. Moreover, the frequency tuning of the satellites with applied DC voltage follows exactly the trend expected from Eq.~\eqref{eq:4wm}. Note that the frequency sum of the satellite resonances always equals the pump frequency $f_3+f_4=f_1+f_2=f_d=2f_c$.\\
As the parametric resonances are tuned closer to each other, at a voltage of $-14.31$\,V corresponding to a frequency difference of $2.41$\,kHz, the two modes suddenly lock to a single, extremely narrow high amplitude line and remain frequency locked for DC voltages exceeding the measurement span. The frequency of the line ($f_5=6.59082$\,MHz) equals exactly half of the pump frequency
\begin{equation}
\label{eq:lock}
f_5=\frac{1}{2}f_d=f_c.
\end{equation}
We attribute this threshold effect to the appearance of injection locking\,\cite{1973_Adler_IEEE,1973_Buczek_IEEE} of the two mechanical modes to half of the common pump frequency. As can be deduced from the extremely large locking range, this state is highly stable against perturbations of the system, in this case against the tuning of the natural resonance frequencies of the two modes. The linewidth narrowing and the high amplitude alongside the large locking range and stability strongly resemble the effect of synchronization of the two modes to the common drive. Since we are not able to simultaneously measure the oscillation phases of the two mechanical resonances in the present experimental set-up, we avoid to denominate the frequency injection locked line as a synchronized state of the two modes. However, the two coupled flexural modes of the nanomechanical string resonator locked to the same frequency\,\cite{2011_Thevenin_PRL} can be considered as a mechanical analog of an electrical quadrature LC oscillator following the work of Mirzaei \textit{et al.}\,\cite{2007_Mirzaei_IEEE}.\\
\begin{figure}[!htb]
\includegraphics{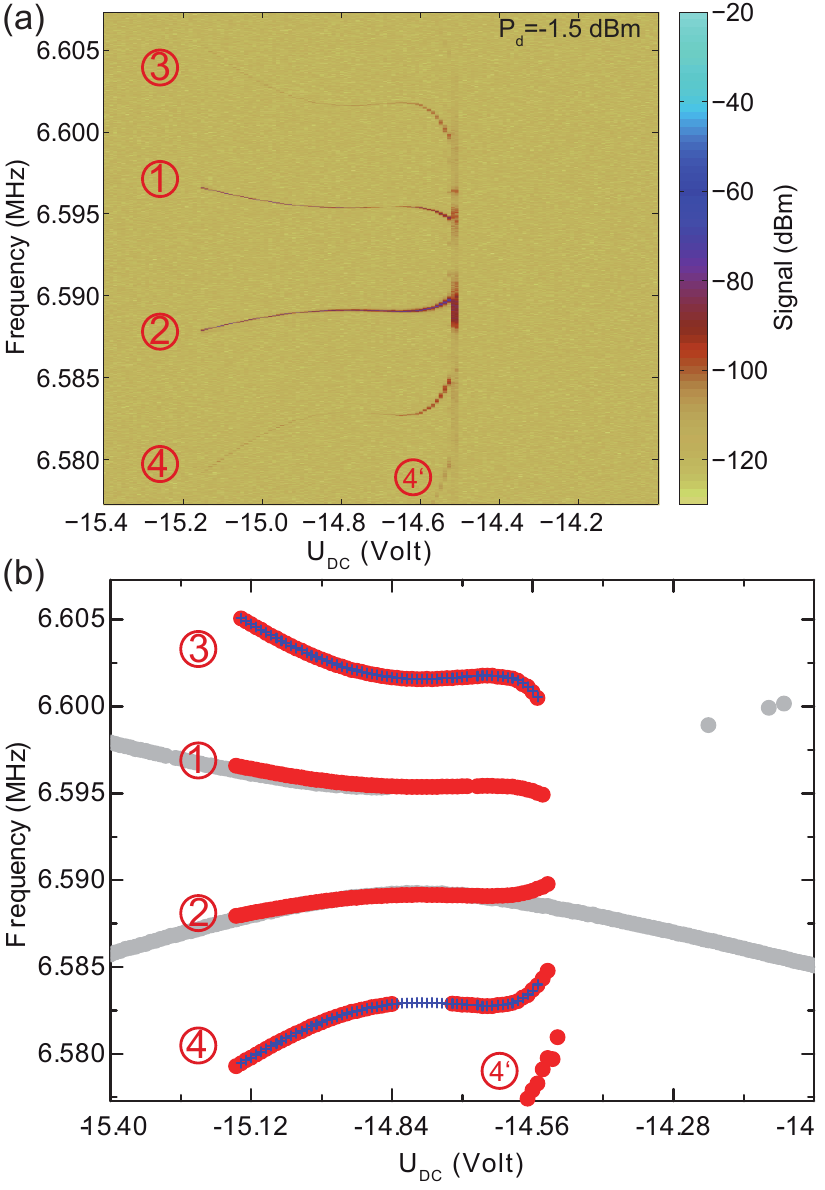}
\caption{\label{fig2}(a) Color-coded frequency spectra versus DC voltage for a weak parametric drive power $P_\mathrm{d}=-1.5$\,dBm at $f_\mathrm{d}=13.18454$\,MHz. Parametric oscillation resonances ($f_1$, $f_2$), satellites from degenerate four-wave-mixing ($f_3$, $f_4$), higher order satellite ($f_{4'}$) and a chaotic regime are observed. (b) Extracted peak frequencies of the above spectra (red dots) and calculated satellite resonances from Eq.~\eqref{eq:4wm}(blue daggers) exhibit excellent agreement. Gray dots display the avoided crossing of the two modes under weak resonant drive with mode splitting of $6.16$\,kHz. Some gray dots are missing for the in-plane mode due to bad detection efficiency of this mode.}
\end{figure}
Figure\,\ref{fig2}\,(a) and Fig.\,\ref{fig2}\,(b) depict the results of the same measurement scheme for a weak parametric drive of power $P_\mathrm{d}=-1.5$\,dBm and a different drive frequency $f_d=13.18454$\,MHz. Again, we observe parametric oscillation ($f_1$, $f_2$) and D4WM ($f_3$, $f_4$) starting at $U_\mathrm{DC}=-15.15$\,V. In addition, another satellite appears at the bottom in Fig.\,\ref{fig2}. The additional satellite has the same spacing from $f_4$ as the first order satellites ($f_3$, $f_4$) from the signal and idler resonance ($f_1$, $f_2$). The symmetric counterpart of this second order satellite resonance is barely visible at the top of Fig.\,\ref{fig2}\,(a) with a too small signal power for the peak frequency extraction of the color-coded data. The appearance of second order satellite resonances can again be interpreted as a D4WM process of the signal-idler frequencies and the first order satellites ($f_{4'}=2f_4-f_2$). Equivalently expressed, the beat-note between signal and idler creates satellite resonances which are off-set by integer multiples of the beat-note.\\
Interestingly, the two parametric resonances do not exhibit injection locking to the pump in this experiment since the parametric drive is not strong enough to overcome the threshold of injection locking. Instead, the signal and idler enter a chaotic regime at $U_\mathrm{DC}=-14.54$\,V and disappear after a short DC voltage range\,\cite{2009_Karabalin_PRB,2013_Bagheri_PRL,2016_Gil-Santos_arxiv}.\\
Note that the coupling strength is $6.16$\,kHz in this measurement although the experiment is conducted on the same sample. Under continuous large-amplitude vibration, a decrease of the coupling strength was observed which we attribute to a reduction of the electrical polarizability of the silicon nitride string as detailed in the Supplemental Material\,\cite{SI_lock}.\\
\begin{figure}[!htb]
\includegraphics{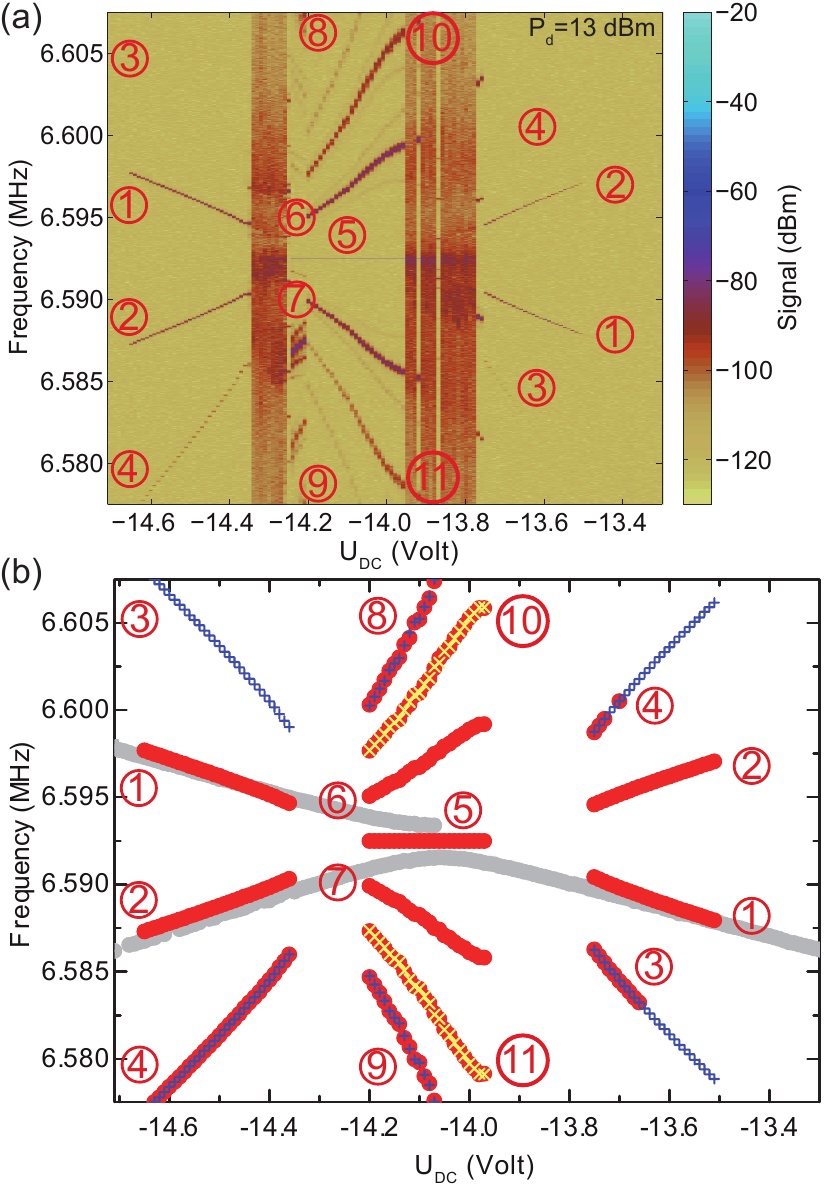}
\caption{\label{fig3}(a) Color-coded frequency spectra versus DC voltage for an intermediate parametric drive power $P_\mathrm{d}=13$\,dBm at $f_\mathrm{d}=13.18502$\,MHz. Parametric oscillation resonances ($f_1$, $f_2$), satellites from degenerate four-wave-mixing ($f_3$, $f_4$) and a partial injection locking peak ($f_5$) are observed. Additionally, equidistant injection pulling peaks appear ($f_6$ to $f_{11}$). (b) Extracted peak frequencies of the above spectra (red dots) and calculated satellite resonances from Eq.~\eqref{eq:mixing}(blue daggers, yellow crosses) exhibit excellent agreement. Gray dots display the avoided crossing of the two modes under weak resonant drive with mode splitting of $1.87$\,kHz. Some gray dots are missing for the in-plane mode due to bad detection efficiency of this mode.}
\end{figure}
In Fig.\,\ref{fig3}\,(a) and Fig.\,\ref{fig3}\,(b) we show the results of the experiment for an intermediate parametric drive power $P_\mathrm{d}=13$\,dBm at a drive frequency $f_\mathrm{d}=13.18502$\,MHz, where the coupling of the modes is extracted as $1.87$\,kHz from the avoided crossing data. Analogous to the above measurements, the system exhibits signal and idler resonances via parametric oscillation and satellite resonances off-set by the beating frequency of signal and idler starting at $U_\mathrm{DC}=-14.65$\,V. When the modes are tuned closer towards each other in frequency, they enter a chaotic regime similar to the results in Fig.\,\ref{fig2}. At a DC voltage of $U_\mathrm{DC}=-14.20$\,V, the modes start to lock to half of the drive frequency $f_5=f_\mathrm{d}/2$ out of the chaotic regime. However, the drive power is apparently not high enough to completely lock the two modes to the common drive. Interestingly, the system exhibits additional resonances compared to the strong pump experiment in Fig.\,\ref{fig1}. The satellite resonances ($f_6$ to $f_{11}$) are equidistantly off-set from the locked resonance by integer multiples of a frequency $f_\mathrm{b}$ which increases with the DC voltage. Following the work of Razavi\,\cite{2004_Razavi_IEEE}, we attribute the appearance of the satellites to the process of injection pulling\,\cite{2004_Razavi_IEEE} in combination with the observed partial injection locking to $f_5$\,\cite{1966_Stover_IEEE,1998_Annovazzi_IEEE,2001_Liu_IEEE,2003_Liu_IEEE,2004_Cross_PRL,2011_Thevenin_PRL}. In the work of Razavi\,\cite{2004_Razavi_IEEE}, the appearance of injection-pulled satellite resonances in a single quasi-locked electrical oscillator was found only on the positive frequency off-set side of the injection locking peak at $f_\mathrm{inj}+n f_\mathrm{b}=f_\mathrm{d}/2+n f_\mathrm{b}\,\,(n=1,2,3,...)$. The frequency off-set from the injection locking frequency $f_\mathrm{inj}$ equals $f_\mathrm{b}=((f_0-f_\mathrm{inj})^2-\delta^2)^{1/2}$, where
\begin{equation}
\label{eq:delta}
\delta\approx \frac{f_0}{2Q} \frac{A_\mathrm{inj}}{A_0}
\end{equation}
represents the maximum locking range which depends on the amplitude ratio of the quasi injection locked oscillation ($A_\mathrm{inj}$) and the oscillation ($A_0$) at the natural resonance frequency $f_0$. In this work, the partial locking of two strongly coupled resonator modes to the common drive leads to additional injection-pulled satellites at $f_\mathrm{d}/2-n f_\mathrm{b}$, the negative frequency off-set side of the injection locking peak. The satellite resonances are completely axial-symmetric with respect to the partial locking resonance.\\
Furthermore, assuming that $f_5$, $f_6$ and $f_7$ are present, the remaining satellites can as well be obtained by frequency mixing
\begin{equation}
\label{eq:mixing}
\begin{aligned}
f_8&=2f_6-f_7
\qquad
f_9=2f_7-f_6\\
f_{10}&=2f_6-f_5
\qquad
f_{11}=2f_7-f_5
\end{aligned}
\end{equation}
depicted as blue daggers and yellow crosses in Fig.\,\ref{fig3}\,(b).\\
By further detuning the natural resonance frequencies from the avoided crossing via the DC voltage, the partial locked and injection pulled state translates into another chaotic regime. At a voltage of $U_\mathrm{DC}=-13.75$\,V the signal and idler resonances reappear out of the chaotic regime and resemble the dynamics of the system on the left hand side of the avoided crossing until the parametric oscillation vanishes at $U_\mathrm{DC}=-13.51$\,V. A comparison of the data provided in Fig.\,\ref{fig3} to the model of Adler\,\cite{1973_Adler_IEEE} is carried out in the Supplemental Material\,\cite{SI_lock}.\\
In conclusion, we investigated the dynamics of two strongly coupled, high quality factor nanomechanical resonator modes in the region of their avoided crossing subject to a common parametric drive. Depending on the driving strength and the frequency spacing of the two modes, we observed signal-idler generation via parametric oscillation in combination with satellite resonances at their beat-note frequency, partial injection locking and two-mode injection pulling, as well as complete injection locking of the two coupled modes to the common drive frequency. By tuning the resonance frequency separation of the two-mode parametric oscillation, we demonstrated control of the nanomechanical signal-idler pair in all examined parametric driving regimes which opens up the path towards parametric control in multimode-nanoelectromechanics. For strong parametric driving, we have been able to completely injection lock the two-mode system to the external parametric pump, a state of extremely narrow mechanical linewidth and high amplitude which is highly stable against perturbations.\\
In future experiments, the presented scheme will be extended in order to study the implications of synchronization of the two strongly coupled modes. Another interesting prospect of the presented scheme is the study of mechanical entanglement\,\cite{2008_Hartmann_PRL,2014_Szorkovszky_NJP,2015_Abdi_NJP} by the parametric degenerate drive. However, this would imply an increase in the mode frequencies to reduce the phonon occupation down to a quantum state as well as drastic improvements of the detection principle towards a quantum back-action-evading measurement\,\cite{2010_Hertzberg_NatPhys,2013_Woolley_PRA,2016_Wollman_Science}.
%
\begin{acknowledgments}
Financial support by the Deutsche Forschungsgemeinschaft via the collaborative research center SFB 767 is gratefully acknowledged. M.A. acknowledges support by the Alexander von Humboldt Foundation via a fellowship for Postdoctoral Researchers. M.J.H. acknowledges support by the Deutsche Forschungsgemeinschaft via the Emmy Noether fellowship HA 5593/1-1. Furthermore, we thank Andreas Isacsson, Yaroslav Blanter and Gianluca Rastelli for fruitful discussions. Preliminary experiments on the topic performed by Onur Basarir and Johannes Rieger are gratefully acknowledged.
\end{acknowledgments}

%
%
%
\renewcommand{\thefigure}{S\arabic{figure}}
 \renewcommand{\theequation}{S\arabic{equation}}
 \renewcommand{\thetable}{S\arabic{table}}

 \renewcommand{\citenumfont}[1]{S#1}
 \renewcommand{\bibnumfmt}[1]{[S#1]}
 
 \setcounter{figure}{0}
 \setcounter{equation}{0}
 
 \clearpage

 \renewcommand{\refname}{Supplemental Material References}

\section*{Supplemental Material to "Parametric oscillation, frequency mixing and injection locking of strongly coupled nanomechanical resonator modes"}

\pagestyle{empty}

\section{Experimental set-up}
\label{sec:set-up}
\begin{figure}[!htb]
\includegraphics{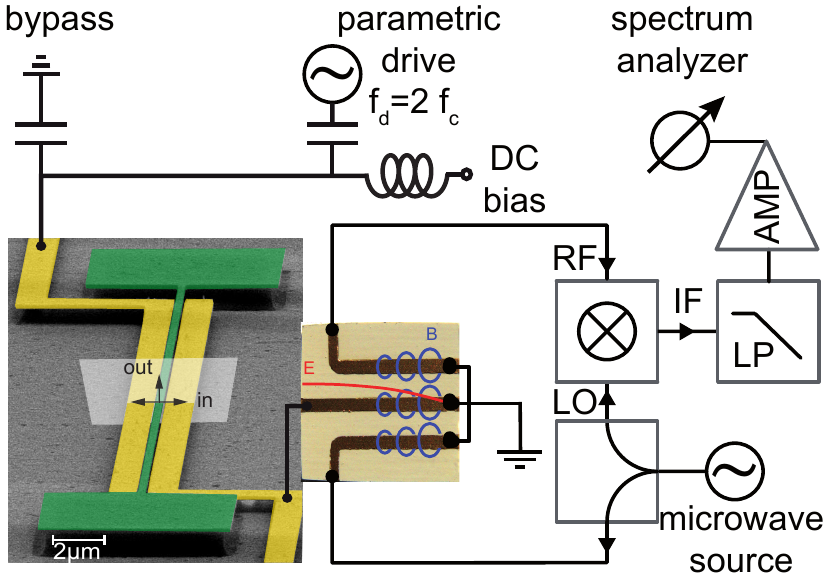}
\caption{\label{fig:setup}Experimental set-up. False-color SEM micrograph of a similar sample depicts the freely suspended, doubly clamped SiN string (green) flanked by the two adjacent gold electrodes (yellow), which are coupled to the $\lambda /4$ microstrip cavity\,\cite{SI_2012_Faust_NatComm}. Red line and blue circles display the electric and magnetic field lines of the cavity, respectively. The transmitted and modulated microwave cavity signal (RF) is demodulated by IQ-mixing with a reference signal (LO). The created intermediate frequency signal (IF) is low-pass filtered (LP), amplified (AMP) and the frequency spectrum is recorded using a spectrum analyzer. A microwave bypass enables applying a DC tuning voltage and a drive tone combined at a bias tee. The sinusoidal drive tone for the parametric drive at $f_\mathrm{d}$ is provided by an arbitrary function generator.}
\end{figure}

The presented nanoelectromechanical system is depicted in Fig.\,\ref{fig:setup} and consists of a doubly clamped silicon nitride nanomechanical string resonator coupled to a $\lambda/4$ microwave cavity via two adjacent gold electrodes\,\cite{SI_2012_Faust_NatComm}. The 55\,\textmu m long, 270\,nm wide and 100\,nm thick mechanical resonator of rectangular cross section exhibits two fundamental flexural modes with orthogonal polarizations either in-plane (in) or out-of-plane (out) with respect to the sample plane. The fabrication induced tensile pre-stress of 1.46\,GPa translates to a high unloaded quality factor of $Q_0\approx 500,000$ at a resonance frequency of $f_\mathrm{0}\approx 6.5$\,MHz for the fundamental out-of-plane flexural mode. The flexural in-plane mode exhibits an approximately 150\,kHz higher fundamental resonance due to the rectangular cross section of the string resonator. A microwave bypass\,\cite{SI_2012_Rieger_APL} allows for the application of RF voltages and enables at the same time control of the inhomogeneous electrical field between the gold electrodes and hence the field gradient by a DC voltage. The applied DC voltage induces electric dipoles in the dielectric silicon nitride string resonator which couple in turn to the electric field gradient, hence resulting in a quadratic tuning of the mechanical frequencies with DC voltage\,\cite{SI_2012_Rieger_APL}. Due to careful engineering of the vertical off-set between the electrodes and the mechanical resonator, we find the in-plane polarized oscillation to decrease quadratically in resonance frequency with the applied DC voltage whereas the out-of-plane polarized mode increases quadratically in frequency. At a DC voltage of approximately $-14$\,V the inherent frequency off-set between the two modes is compensated and we resolve a clear avoided level crossing between the two frequency branches. The strong coupling between the two fundamental flexural modes is mediated by the inhomogeneous electric field in terms of non-vanishing cross derivatives of the electric field energy\,\cite{SI_2012_Faust_PRL} as detailed in section\,\ref{sec:coupling}. In order to drive the mechanical oscillation, a RF drive tone is added to the DC voltage using a bias tee. The oscillation of the nanomechanical resonator is detected in a microwave cavity assisted heterodyne in-phase quadrature mixing technique\,\cite{SI_2012_Faust_NatComm}. Hereby, the oscillation of the dielectric silicon nitride string resonator periodically modulates the capacitance of the gold electrodes, generating sidebands in the microwave cavity signal. Note that we operate in the unresolved sideband regime due to a cavity quality factor of $Q_\mathrm{cav}\approx 100$ at a cavity resonance frequency of $f_\mathrm{cav}\approx 3.6$\,GHz. The modulation induced sidebands of the cavity signal are low-pass filtered, amplified and displayed as mechanical frequency spectrum by a spectrum analyzer using a Fast-Fourier-Transform (FFT) Filter.\\
\section{Influence of the drive power on the coupling strength}
\label{sec:coupling}
\begin{figure}[!htb]
\includegraphics{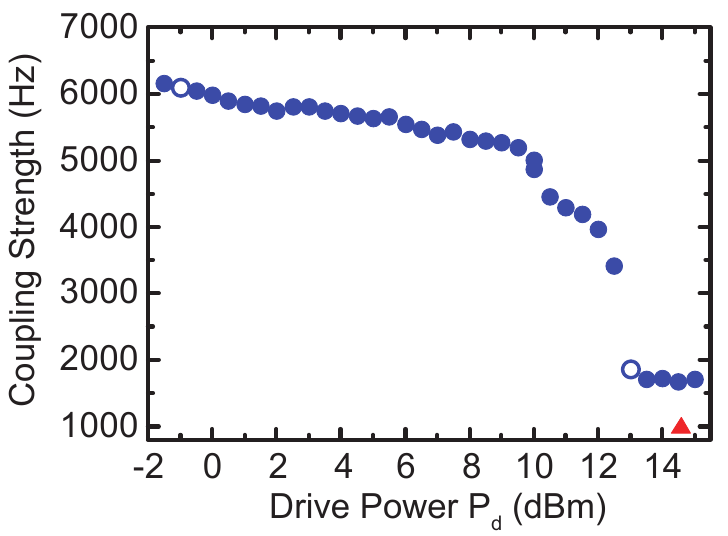}
\caption{\label{coupling} Coupling strength as extracted from the avoided crossing frequency splitting for each particular drive power $P_\mathrm{d}$ (blue dots). The red triangle corresponds to the coupling strength in the measurement of Fig.\,1 of the main text at a drive power $P_\mathrm{d}=14.6$\,dBm. The open circles indicate the datasets displayed in Fig.\,2 and Fig.\,3 of the main text, respectively.}
\end{figure}

The strong coupling of the two nanomechanical flexural modes is provided by the inhomogeneous electric field in between the two gold electrodes. The coupling strength is equivalent to the frequency splitting of the two normal modes at the avoided crossing and can be expressed as\,\cite{SI_2012_Faust_PRL}
\begin{equation}
\label{eq:coupling}
\frac{\Delta}{2\pi}=\frac{1}{2\pi}\left(\sqrt{\frac{k_0+2k_\mathrm{c}}{m_\mathrm{eff}}}-\sqrt{\frac{k_0}{m_\mathrm{eff}}}\right),
\end{equation}
where $k_0$ is the spring constant of the normal modes and $k_\mathrm{c}$ is an asymmetric coupling spring. The coupling spring constant is determined by the cross derivatives of the electric energy $E^2$ with respect to the out-of-plane (z-direction) and the in-plane (y-direction) oscillation of the nanomechanical string\,\cite{SI_2012_Faust_PRL}
\begin{equation}
\label{eq:kc}
k_\mathrm{c}=\alpha \frac{\partial^2 E^2}{\partial z \partial y}.
\end{equation}
Here, $\alpha$ is the polarizability of the silicon nitride string as determined from the electric force acting on it in dipole approximation
\begin{equation}
\label{eq:fel}
\vec{F}_\mathrm{el}= - \vec{\nabla}(\vec{p}\cdot\vec{E}),
\end{equation}
where $\vec{p}=\alpha \vec{E}$ denotes the electric dipole moment.\\
As stated in the main text, the coupling strength of the two modes decreases with increasing drive power of the strong parametric drive. In Fig.\,\ref{coupling} we display the coupling strength extracted from the avoided crossings of the two modes prior to each measurement at different drive powers $P_\mathrm{d}$. Note that we started the measurements at a parametric drive power of $P_\mathrm{d}=-1.5$\,dBm and systematically increased the drive power up to $P_\mathrm{d}=15$\,dBm in steps of $0.5$\,dBm.
We observe a non-reversible, monotonous decrease of the coupling strength as a function of the parametric drive power. However, the coupling does not decrease in a clear functional way. Up to a drive power of $P_\mathrm{d}=9.5$\,dBm, one may qualitatively extract a linear relation between coupling strength decrease and the logarithmically increasing drive power. For larger drive powers, the decrease does not qualitatively follow a simple function.\\
In addition, the coupling strength also decreases if the strongly parametrically driven measurement is repeated at a specific drive power. As can be seen from Fig.\,\ref{coupling}, the red triangle, which corresponds to $P_\mathrm{d}=14.6$\,dBm, depicts a smaller coupling than the blue dot at $P_\mathrm{d}=14.5$\,dBm since this data point was taken after multiple measurements in the region of $14.5$\,dBm$\leq P_\mathrm{d} \leq$$15$\,dBm. Obviously, strong parametric driving of the two coupled modes results in a non-reversible decrease of the electrically mediated coupling.
Surprisingly, the effect can be completely repealed by the exposition of the sample to air. After venting the vacuum chamber and subsequent evacuation of the chamber back to the operation pressure, we reversibly recover the initial electrically mediated coupling strength of the two nanomechanical modes.\\
One possible explanation of this effect may be the implantation of static dipoles in the silicon nitride string due to the high electric fields the dielectric material is exposed to under the strong parametric drive. Static dipoles in the material would decrease the polarizability of the silicon nitride string and hence reduce the electric coupling spring according to Eq.\,\eqref{eq:kc}. When exposed to air, the material would recover from the static dipoles due to charge exchange with the surrounding air molecules. However, the proposed mechanism does apparently not apply straightforwardly to the decrease of the coupling for repeated measurements at a fixed parametric drive power. In a lumped model, one could argue that large oscillation amplitudes of the dielectric string in combination with the high electric fields result in a big electric field gradient, which, in turn, favors the implantation of static dipoles in the material. In the case of the injection locked state, we expect mechanical amplitudes of tens of nanometers from the signal heights in the frequency spectra compared to the signal heights of the resonantly driven modes in the linear regime (typical amplitudes of few hundred picometers). The exact origin of the decreasing mode coupling is currently under investigation and remains subject of future research.
\section{Comparison of the experimental results to the Adler equation}
\label{sec:Adler}
The Adler equation\,\cite{SI_1973_Adler_IEEE} describes the temporal evolution of the frequency and the phase of an oscillator which is impressed with an external signal of similar fundamental frequency. The solutions to this differential equation are periodic waves with a frequency off-set from the injected signal by\,\cite{SI_1973_Adler_IEEE,SI_2014_DelHaye_PRL}
\begin{equation}
\label{eq:Adler}
f_\mathrm{off}=(f_0-f_\mathrm{inj}) \frac{\sqrt{\left(2 Q \frac{A_0}{A_\mathrm{inj}} \right)^2 \left(\frac{f_0-f_\mathrm{inj}}{f_0}\right)^2 -1}}{2 Q\frac{A_0}{A_\mathrm{inj}}\abs{\frac{f_0-f_\mathrm{inj}}{f_0}}}.
\end{equation}
As defined in the main text, $f_0$ and $f_\mathrm{inj}$ are the frequencies of the free-running oscillator and the injected signal, $Q$ is the mechanical quality factor and $A_0$ and $A_\mathrm{inj}$ denote the amplitude of the oscillator and the injected signal, respectively. Note that Eq.~\eqref{eq:Adler} considers only a single mode subject to the parametric drive. Since the signal and idler resonances are in the present experiment symmetrically off-set from the injected signal, it is well justified to consider one of the two resonances exemplarily.\\
If the argument of the square root in Eq.~\eqref{eq:Adler} becomes negative, i.e.,
\begin{equation}
\label{eq:lock_si}
\frac{1}{2 Q} \frac{A_\mathrm{inj}}{A_0} > \abs{\frac{f_0-f_\mathrm{inj}}{f_0}}
\end{equation}
the free-running oscillator locks to the external signal. Hence, the real part of Eq.~\eqref{eq:Adler} becomes zero, which is the frequency off-set from the injected frequency.
\begin{figure}[!htb]
\includegraphics{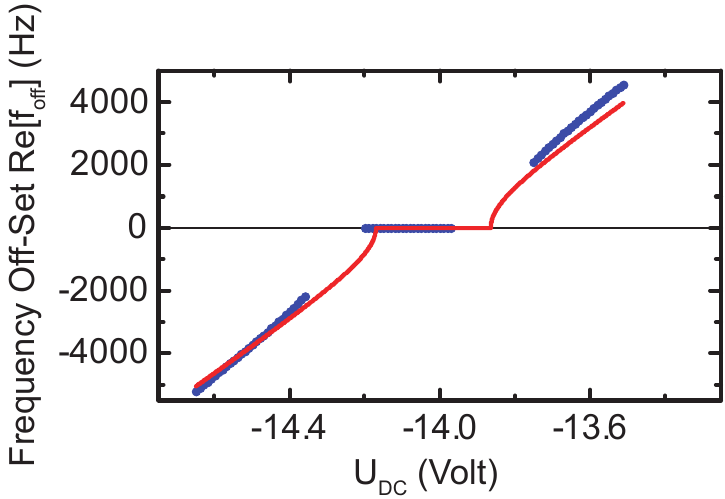}
\caption{\label{adler}Frequency off-set $f_2-f_5$ extracted from Fig.\,3\,(b) of the main text as a function of DC voltage (blue dots). The red line is the fit of the real part of Eq.~\eqref{eq:Adler} using one open parameter.}
\end{figure}
In Fig.\,\ref{adler}, we plot the frequency difference of the parametric oscillation resonance $f_2$ from half of the drive frequency $f_2-f_\mathrm{d}/2=f_2-f_5$ extracted from Fig.\,3\,(b). Note that the missing data points correspond to the regions where the system behaves chaotically. In order to fit Eq.~\eqref{eq:Adler} to the experimental data with a single fit parameter, we apply several assumptions and simplifications. First, we linearize the frequency tuning of the fundamental free-running oscillation supposing $f_0(U_\mathrm{DC})=f_0+c\cdot U_\mathrm{DC}$ in the vicinity of the avoided crossing and extract a linear tuning coefficient $c=8.2$\,kHz/V. Second, the mechanical quality factor is fixed at a value of $Q=200,000$ which is a reasonable value for the employed DC voltage range\,\cite{SI_2016_Seitner_PRB}. Third, the only open fit parameter is the ratio of the amplitudes of free-running and injected signal $A_\mathrm{inj}/A_0$. The fit of Eq.~\eqref{eq:Adler} is displayed as red line in Fig.\,\ref{adler} from which we obtain $A_\mathrm{inj}/A_0=(76.329\pm 0.008)$, where the provided error corresponds to the confidence interval of the fit. As expected, the quantitative agreement of the experimental data and the fit is rather poor due to the drastic approximations and the missing data points in the chaotic regions. Nevertheless, the dynamics of the system qualitatively follow the theoretically predicted locking phenomena.\\
In addition, the injection pulling satellite frequency off-set, as described in the main text and defined by Razavi\,\cite{SI_2004_Razavi_IEEE}, can be recovered by taking the imaginary part of Eq.~\eqref{eq:Adler}:
\begin{equation}
\label{eq:pull}
f_b= \operatorname{Im}[f_\mathrm{off}]=\operatorname{Im}\left[\sqrt{(f_0-f_\mathrm{inj})^2-\left(\frac{f_0}{2 Q} \frac{A_\mathrm{inj}}{A_0}\right)^2}\right].
\end{equation}
Outside of the locking region, Eq.~\eqref{eq:pull} is equal to zero and hence no injection pulling is observed experimentally. Inside of the partial locking region, we observe injection pulling satellites which are equidistantly off-set from the locking peak (cf. Eq.~(5) of the main text). However, the branching of the experimentally observed injection pulling satellite frequencies does not follow the trend calculated from Eq.~\eqref{eq:pull} using the parameters as defined above.

\section*{Supplemental Material References}

\end{document}